\documentclass[journal, 10pt]{IEEEtran}
\usepackage{amssymb}
\usepackage{amsmath, tabularx}
\usepackage{mathtools}
\usepackage{ellipsis}
\usepackage{textcomp}
\usepackage{breqn}
\usepackage{comment}
\usepackage{cite}
\usepackage{graphicx}	%[pdftex]
\usepackage{subfigure}
\usepackage{epsfig}
\usepackage{caption}
\usepackage{psfrag}
\usepackage{grffile}

\usepackage{setspace}
\usepackage{epstopdf} %converting to PDF
\usepackage{color}
\usepackage{multirow}
\usepackage{amsthm}

\usepackage[overload]{empheq}
\usepackage{cases}
\usepackage{url}

\usepackage{pifont}% http://ctan.org/pkg/pifont
\newcommand{\cmark}{\ding{51}}%
\newcommand{\xmark}{\ding{55}}%

\usepackage{xcolor}
\usepackage[noend]{algpseudocode}
\usepackage{algorithm}
\algnewcommand{\LineComment}[1]{\Statex \(\triangleright\) #1} 
\algnewcommand{\And}{\textbf{and}}
\algnewcommand{\Or}{\textbf{or}}
\usepackage{multicol}

\algnewcommand{\LineCommentCont}[1]{\Statex \hskip\ALG@thistlm%
	\parbox[t]{\dimexpr\linewidth-\ALG@thistlm}{\hangindent=\trianglerightwidth \hangafter=1 \strut$\triangleright$ #1\strut}}

\DeclareTextFontCommand{\textOF}{\fontfamily{lmtt}\selectfont}

\IEEEoverridecommandlockouts
\begin{document}

	%	\title{Softwarized Network Security Functions: Implementation, Performance Evaluation, and }
	\title{Security-as-a-Function for IDS/IPS in Softwarized Network and Applications to 5G Network Systems} %Studying the Impact of Software Security Functions on Network Performance
	%\author{Work in Progress}
	\author{Shivank~Malik~and~Samaresh~Bera,~\IEEEmembership{Senior~Member,~IEEE}
		%		\IEEEauthorblockA{Department of Computer Science and Engineering\\
			%			Indian Institute of Technology Jammu, 181221, India\\
			%			Email: 2022pis0074@iitjammu.ac.in
			%			}
		%		\and
		%		\IEEEauthorblockN{Samaresh~Bera,~\textit{Member,~IEEE}}
		%		\IEEEauthorblockA{Department of Computer Science and Engineering\\
			%				Indian Institute of Technology Jammu, 181221, India\\
			%				Email: s.bera.1989@ieee.org
			%			}
		\thanks{S.~Malik and S.~Bera are with the Department of Computer Science and Engineering, Indian Institute of Technology Jammu, Jammu and Kashmir, India, 181221. Email: \{2022pis0074@iitjammu.ac.in, s.bera.1989@ieee.org\}.}
		\thanks{A preliminary version~\cite{shivankSecurityasaFunction5GNetwork2024} of this work has been accepted for publication in the International Conference on Signal Processing and Communications (SPCOM), 2024, Bangalore, India.}
	}
	
	%\DeclareGraphicsExtensions{.pdf,.jpg,.bmp,.gif,.eps}
	\maketitle
	
	%	\doublespacing
	\begin{abstract}
		
		The service-based architecture of 5G network allows network operators to place virtualized network functions on commodity hardware, unlike the traditional vendor-specific hardware-based functionalities. However, it expands the security vulnerabilities/threats to the 5G network. While there exist several theoretical studies on network function placement and service routing, a few focused on the security aspects of the 5G network systems.
		
		This paper focuses on safeguarding the 5G core network systems from DoS/DDoS attacks by placing intrusion detection/prevention systems (IDS/IPS) as virtualized network functions following the 5G standalone architecture. To ensure the virtualized placement of IDS/IPS, first, we provide thorough virtual machine (VM)-based and containerized implementation details and evaluate the network performance with two scenarios – IDS and IPS – in the presence of TCP and UDP applications. Second, we apply the VM-based implementation of IDS/IPS on a softwarized 5G core network and study the network performances. The experiment results on network throughput, latency, and packet drop reveal that the softwarized IDS/IPS can meet the QoS requirements of 5G applications, while safeguarding the network from DoS/DDoS attacks.
	\end{abstract}
	
	
	%================================Introduction====================================
	\section{Introduction} \label{Sec:Introduction}
	
	% background about softwarized networks and VNFs using SDN and NFV
	
	The recent advancement of mobile communications (such as 5G and beyond networks) and industry 5.0 is expected to support modern applications with stringent quality-of-service (QoS) requirements. The 5G applications are categorized into three broad areas -- enhanced mobile broadband (eMBB), ultra-reliable and low-latency communications (uRLLC), and massive machine-type communications (mMTC)~\cite{NavarroOrtizSurvey5GUsage2020, VerticalsURLLCUse2020a}. The QoS requirements of these applications range from high bandwidth to high reliability to low latency~\cite{5GServiceRequirements}. Whereas Industry 5.0 requires network operators to enable smart manufacturing with improved resilience and sustainability~\cite{xianAdvancedManufacturingIndustry2023}. Furthermore, it also requires the network operators to support a massive number of smart devices (such as robots) in a small and congested deployment area. Consequently, heterogeneous applications need to be supported by the network while preserving security to the network elements. The ultimate choice is to place closed-box vendor-specific hardware devices to ensure a secure system. However, fulfilling such diverse QoS requirements of the applications using traditional vendor-specific networking architecture and devices are complex and cost-expensive~\cite{jarrayaSurveyLayeredTaxonomy2014}.	
	
	The softwarized network supported by software-defined networking (SDN) and network function virtualization (NFV) technologies addresses the issues with vendor-specific networking devices~\cite{nunesSurveySoftwaredefinedNetworking2014, gilherreraNetworkFunctionsVirtualization2016, farhadySoftwaredefinedNetworkingSurvey2015}. The SDN provides flexible networking for traffic forwarding by separating the control and data planes. Whereas NFV enables network operators to place virtual network functions (VNFs) on commodity hardware as per requirements. Therefore, with the help of SDN and NFV, customized network functions can be placed as VNFs on commodity hardware in the form of virtual machines (VMs) or containers~\cite{gilherreraNetworkFunctionsVirtualization2016, gediaPerformanceEvaluationSDNVNF2018, cerroniNetworkSoftwarizationManagement2020}.
	
	The service-based architecture of 5G network network also allows the use of virtualized network functions instead of closed-box hardware devices~\cite{3gpp5GSystemOverview}. Furthermore, there exist significant studies on the placement of VNFs, traffic scheduling through the VNFs, and efficient resource utilization at the commodity hardware~\cite{agarwalJointVNFPlacement2018, liuJointDynamicalVNF2021, guptaPerformanceAnalysis2020, salahPerformanceComparisonContainerbased2017, wangCanMicroVM2021}. However, these studies focused on theoretical modeling of the problem and did not focus on the practical implementation and their impact on network performance, i.e., whether the QoS requirements of the underlying applications can be fulfilled using softwarized functions. Moreover, the use of softwarized network expands the security threats/vulnerabilities compared to the traditional hardware-based networking. Furthermore, the network performance may be impacted with softwarized functions compared to the hardware-based systems. While there exist a few works on studying the impact on network performances~\cite{gallenmuller5GQoSImpact2020, vargheseEfficientIDSFramework2021, shamsComparativeAnalysisIntrusion2023, steinbergerHowExchangeSecurity2015, fadhilahPerformanceAnalysisIDS2020, seeberSDNenabledIDSEnvironment2015}, a comprehensive study is required considering emerging applications with stringent QoS requirements, as presented in Table~\ref{table:5G3GPP}~\cite{5GServiceRequirements}, to analyze the feasibility of using softwarized network functions in 5G network system while ensuring security.
	
	\begin{table*}[!ht]
		\centering
		\caption{QoS parameters of different applications~\cite{5GServiceRequirements}}
		\label{table:5G3GPP}
		\begin{tabular}{cccccc}
			\hline
			\multicolumn{2}{c}{\textbf{Scenario}}                                                                                                                   & \textbf{Latency} & \textbf{Data rate} & \textbf{Reliability} & \textbf{Payload} \\ \hline
			\multicolumn{2}{c}{Discrete automation}                                                                                                                 & 10 ms            & 10 Mbps            & 99.99\%                & small to high    \\ \hline
			\multirow{2}{*}{\begin{tabular}[c]{@{}c@{}}Process\\ automation\end{tabular}}       & remote control                                                    & 60 ms            & 1 to 100 Mbps      & 99.999\%               & small to high    \\ \cline{2-6} 
			& monitoring                                                        & 60 ms            & 1 Mbps             & 99.9\%                 & small            \\ \hline
			\multirow{2}{*}{\begin{tabular}[c]{@{}c@{}}Electricity\\ distribution\end{tabular}} & medium voltage                                                    & 40 ms            & 10 Mbps            & 99.9\%                 & small to high    \\ \cline{2-6} 
			& high voltage                                                      & 5 ms             & 10 Mbps            & 99.999\%               & small            \\ \hline
			\begin{tabular}[c]{@{}c@{}}Intelligent\\ transportation system\end{tabular}         & \begin{tabular}[c]{@{}c@{}}infrastructure\\ backhaul\end{tabular} & 30 ms            & 10 Mbps            & 99.9999\%              & small to high    \\ \hline
		\end{tabular}
	\end{table*}
	% \usepackage{multirow}
	\begin{table*}
		
		\centering
		\caption{Comparison between the existing studies and this work}
		% \usepackage{multirow}
		\begin{tabular}{l|ccc|cccc|cc|c}
			\hline
			\multicolumn{1}{c|}{\multirow{2}{*}{\textbf{Work}}} & \multicolumn{3}{c|}{\textbf{Security Functions}} & \multicolumn{4}{c|}{\textbf{Performance Analysis}}                               & \multicolumn{2}{c|}{\textbf{Environment}} & \multirow{2}{*}{\textbf{\begin{tabular}[c]{@{}c@{}}Use\\ Case\end{tabular}}} \\ \cline{2-10}
			\multicolumn{1}{c|}{}                               & \textbf{IDS}   & \textbf{IPS}   & \textbf{NAT}   & \textbf{Thr.} & \textbf{Lat.} & \textbf{Jit.} & \textbf{PD} & \textbf{VM}     & \textbf{Containers}     &                                                                              \\ \hline
			Gallenmuller et al.~\cite{gallenmuller5GQoSImpact2020}                                 & \cmark            & \cmark              & \xmark              & \xmark                   & \cmark                & \xmark               & \xmark                     & \cmark               & \xmark  &     \xmark                 \\ \hline
			Seeber et al.~\cite{seeberSDNenabledIDSEnvironment2015}                                       & \cmark              & \xmark              & \xmark              & \xmark                   & \xmark                & \xmark               & \xmark                     & -               & -          &     \xmark         \\ \hline
			Fadhilah and Marzuki~\cite{fadhilahPerformanceAnalysisIDS2020}                                & \cmark              & \xmark              & \xmark              & \xmark                   & \xmark                & \xmark               & \cmark                     & \cmark               & \xmark        & \xmark               \\ \hline
			Aggarwal and Thangaraju~\cite{aggarwalPerformanceAnalysisVirtualisation2020}                             & \xmark              & \xmark              & \xmark              & \cmark                   & \xmark                & \xmark               & \xmark                     & \cmark               & \cmark        & \xmark               \\ \hline
			Li et al.~\cite{liPerformanceOverheadComparison2017}                                           & \xmark              & \xmark              & \xmark              & \cmark                   & \xmark                & \xmark               & \xmark                     & \cmark               & \cmark         & \xmark              \\ \hline
			Garg et al.~\cite{gargMigratingVMWorkloads2018}                                         & \cmark              & \xmark              & \xmark              & \xmark                   & \xmark                & \xmark               & \xmark                     & \xmark               & \cmark         & \xmark              \\ \hline
			Gupta and Sharma~\cite{guptaPerformanceAnalysis2020}                                    & \cmark              & \xmark              & \xmark              & \xmark                   & \xmark                & \xmark               & \cmark                     & \cmark               & \xmark          & \xmark             \\ \hline
			Varghese and Muniyal~\cite{vargheseEfficientIDSFramework2021}                                & \cmark              & \xmark              & \xmark              & \cmark                   & \cmark                & \xmark               & \xmark                     & \cmark               & \cmark            & \xmark           \\ \hline
			\textbf{This work}                                  & \cmark              & \cmark              & \cmark              & \cmark                   & \cmark                & \cmark               & \cmark                     & \cmark               & \cmark        & \cmark               \\ \hline
		\end{tabular}
		\label{table:comparison}
	\end{table*}

	In this paper, we study the implementation and impact of softwarized network security functions on network performance in a softwarized 5G network. In particular, we focus on the primary network security functions, such as intrusion detection system (IDS) and intrusion prevention system (IPS). We note that we utilize both the containerized and VM-based implementations of the security functions on commodity hardware. The objective of performing this study in both VM and containerized environment is to obtain and analyze a holistic overview of the security functions on network performance and not to compare them with physical network functions. Next, we apply the softwarized IDS/IPS functions in a softwarized 5G network to safeguard the other network functions in the 5G core network from DoS/DDoS attacks. We also study the network performance in the 5G network in terms of throughput, latency, and packet drop. The key contributions in this work are as follows:
	
	\begin{itemize}
		\item We implement a secure softwarized network with IDS and IPS virtual network functions as VMs and containers placed on commodity hardware. Furthermore, we utilize open-source software tools to implement the network security functions.
		
		\item We provide comprehensive details of the network setup for two implementation scenarios, \textOF{IDS} and \textOF{IPS}, for both VM-based and containerized placement. 
		
		\item We use the D-ITG~\cite{ditg} traffic generator to generate network traffic for TCP and UDP applications with varying QoS requirements. The extensive experiment results on throughput, latency, jitter, and packet drop are analyzed considering diverse QoS requirements of different applications, as presented in Table~\ref{table:5G3GPP}.
		
		\item We apply the network security functions in a softwarized 5G network and build a prototype of the entire system. Then we present the results on network throughput, latency, and packet drop in the presence of the security functions. The results show that the 5G network can meet the QoS requirements of emerging 5G applications, while safeguarding it from the DoS/DDoS attacks.
	\end{itemize}

	The remainder of the paper is organized as follows. Section~\ref{Sec:Related_work} presents an overview of the existing studies. Section~\ref{Sec:Implementation} describes the detailed network setup and implementation of a secure softwarized network. Section~\ref{Sec:Results_discussion} presents the experiment results with a detailed analysis. Section~\ref{Sec:Use_case_5G} presents the 5G network prototype with security functions and results on network performance. Finally, Section~\ref{Sec:Conclusion} concludes the paper with future research directions.

	%=======================================================
	\section{Background}\label{Sec:Related_work}
	In this section, we discuss the exiting studies on IDS/IPS  and their impacts on network performance~\cite{5GServiceRequirements, gediaPerformanceEvaluationSDNVNF2018, liPerformanceOverheadComparison2017, guptaPerformanceAnalysis2020, salahPerformanceComparisonContainerbased2017, wangCanMicroVM2021, gallenmuller5GQoSImpact2020, vargheseEfficientIDSFramework2021, shamsComparativeAnalysisIntrusion2023, steinbergerHowExchangeSecurity2015, fadhilahPerformanceAnalysisIDS2020, seeberSDNenabledIDSEnvironment2015, kellererHowMeasureNetwork2018}.
	
	Fadhilah and Marzuki~\cite{fadhilahPerformanceAnalysisIDS2020} discussed the impact of hardware configuration of VMs using different numbers of CPU cores while creating different intrusion-based attack scenarios in the network. In another study, Gallenmuller et al.~\cite{gallenmuller5GQoSImpact2020} evaluated the impact of Snort-based forwarding and filtering strategies on latency performance. The authors evaluated the performance of the Snort-based security functions in two scenarios -- functions placed on the host machine and VMs. The results show that VM-based security functions impose increased latency compared to the direct placement on the host machine.
	
	Shams et al.~\cite{shamsComparativeAnalysisIntrusion2023} studied the comparison of different security functions in a software-defined networking platform. The authors integrated an open-source IDS software with SDN and tested their capabilities against malicious traffic generated by an attacker. Varghese and Muniyal~\cite{vargheseEfficientIDSFramework2021} proposed a framework for distributed denial-of-service (DDoS) attacks in SDN environment. The attack report captured at the data-plane of the SDN is sent to the control-plane for making fine-grained traffic forwarding decisions.
	
	Gedia and Perigo~\cite{gediaPerformanceEvaluationSDNVNF2018} presented SDN controller's performance when implemented inside a VM and a containerized platform. In another study, Li et al.~\cite{aggarwalPerformanceAnalysisVirtualisation2020} analyzed the performance of various virtualization technologies for VNF deployments, highlighting trade-offs between isolation and agility. These studies majorly delve into the performance implications of NFVs. Similarly, the works in~\cite{salahPerformanceComparisonContainerbased2017, wangCanMicroVM2021, guptaPerformanceAnalysis2020} explored the placement and resource utilization of VNFs, highlighting their potential for efficient service provisioning.
	
	The authors in~\cite{liPerformanceOverheadComparison2017, felterUpdatedPerformanceComparison2015} presented a comparison between VM and container overhead, emphasizing the need for careful tuning, especially for I/O-intensive security functions such as IDS and IPS. Beyond direct performance comparisons, Garg et al.~\cite{gargMigratingVMWorkloads2018} examined the challenges associated with migrating VM-based workloads (e.g., security functions) to containers. The study considers resource sharing, concurrency, isolation and dependability, while providing valuable insights into potential performance implications of such migrations. 
	
	While some approaches offer potential benefits, a careful consideration of the platform, configuration of security functions and its characteristics are crucial to ensure QoS-aware service provisioning in practical deployments. Furthermore, while a few are aligned with the studied scenarios in this paper, they significantly differ from this work. The key differences are highlighted briefly in Table~\ref{table:comparison} and discussed as follows. This work focuses on a comprehensive implementation and performance evaluation of softwarized security functions. While most of the existing works focus on the detection and prevention of intrusions, our objective is to study the impact of softwarized network security functions on network performance. Therefore, we do not focus on the decision making policies on traffic forwarding, i.e., whether to forward or drop or generate alert message upon receiving an incoming network traffic. The evaluation is also done in a softwarized 5G network as a use case scenario in contrast to the existing works.

	%======================================================
	%\section{Framework for Studying Network Performance}\label{Sec:Framework}
	
	%========================================================
	\section{Implementation Details}\label{Sec:Implementation}
\begin{figure}[!ht]
		\centering
		\includegraphics[scale=0.5]{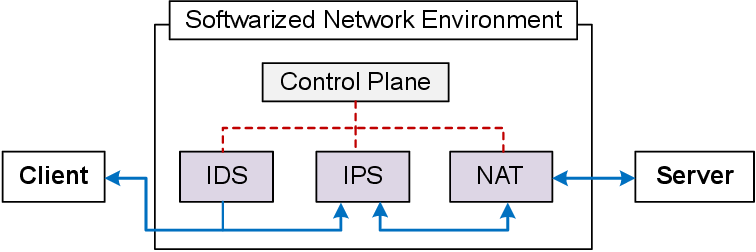}
		\caption{Schematic view of the framework considered for softwarized security function}
		\label{fig:framework}
	\end{figure}

	We follow the ETSI's NFV architecture~\cite{NFVReleaseDescription2021} to deploy and the manage the VNFs in a virtualized environment. Consequently, we use the virtualized infrastructure of a host machine to place the security functions with the following hardware and software configuration: Processor: Intel i9-13th generation, 24 core; RAM: 64GB; Ethernet: 1 Gbps.
	%	\begin{figure}[!ht]
		%		\centering
		%		\includegraphics[scale=0.6]{Images/NFV_Architecture}
		%		\caption{ETSI's NFV architecture for placement of VNFs~\cite{NFVReleaseDescription2021}}
		%		\label{fig:NFV_architecture}
		%	\end{figure}
	Figure~\ref{fig:framework} shows the schematic view of the security framework between the client and server. The network traffic between the client and server is passed through the secure virtualized network based on the security configurations. We consider the following network security functions -- IDS and IPS, as shown in Figure~\ref{fig:framework}. We note that the network address translation (NAT) module is the integral part of the network system.
	
\begin{figure}[!ht]
		\centering
		\subfigure[IDS mode]{
			\includegraphics[scale=0.55]{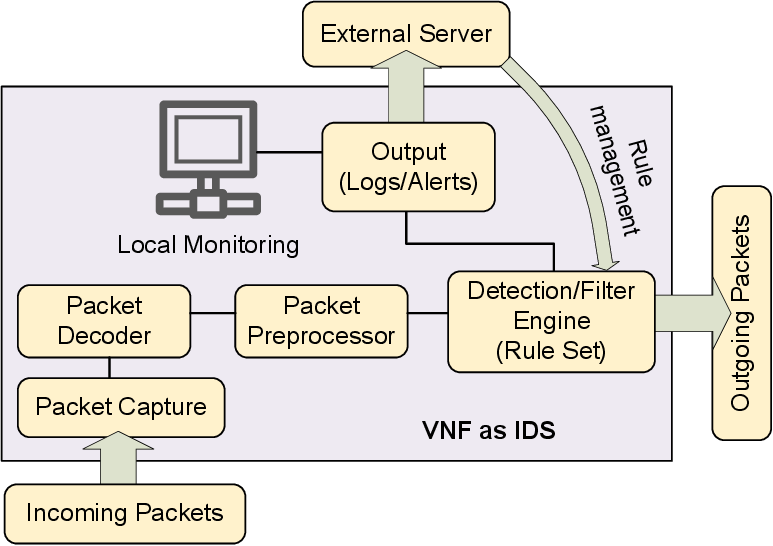}
			\label{fig:VNF_IDS}
		}
		\subfigure[IPS mode]{
			\includegraphics[scale=0.55]{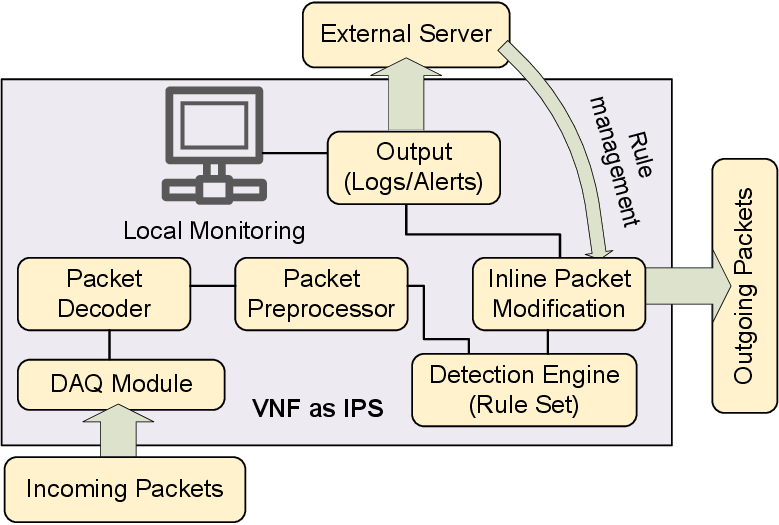}
			\label{fig:VNF_IPS}
		}
		\caption{Components of the network security function in IDS and IPS modes}
		\label{fig:VNF_architecture}
	\end{figure}

	\textbf{Network Setup:} We create three guest machines on the host machine, two for the client and the server, and the other to place the softwarized security functions. Similarly, three containers are created to understand the impact of softwarized security functions in containerized environment. We use Snort~(\url{https://www.snort.org/}), which is an open-source network security tool, to create the IDS and IPS network functions. In case of IPS, we use the data acquisition library (DAQ)~(\url{https://github.com/snort3/libdaq}) in order to efficiently intercept, analyze and take rule-based actions on the network traffic. At the same time, we use the \textOF{iptables} utility available in Linux to create NAT forwarding functionalities. Figures~\ref{fig:VNF_IDS} and \ref{fig:VNF_IPS} present the modules of the VNF in IDS and IPS modes, respectively. Consequently, we create different security scenarios discussed in the subsequent sections.
	
	%---------------------------------------------------------------------------
	\subsection{Scenario 1: Intrusion Detection System (IDS)}\label{Secsub:scenario_ids_nat}
\begin{figure}[!ht]
		\centering
		\includegraphics[scale=0.5]{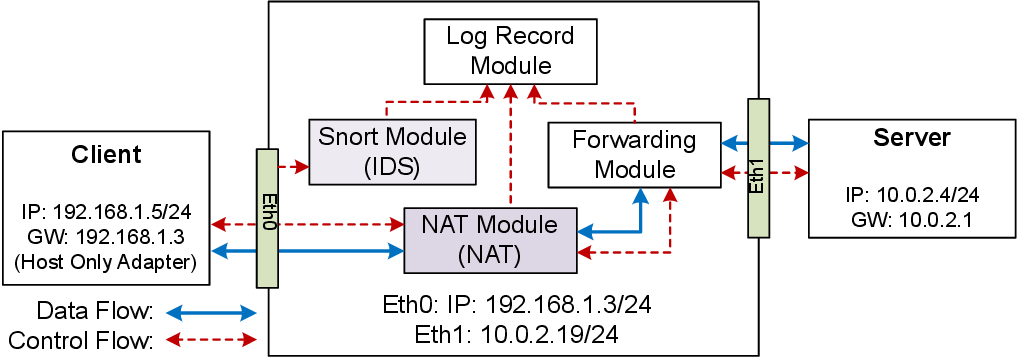}
		\caption{IDS between client and server}
		\label{fig:implementation_scenario_ids_nat}
	\end{figure}

	Figure~\ref{fig:implementation_scenario_ids_nat} shows the network setup between the client and server with IDS software security functions. The IP configuration of client, server, and virtualized network is shown in Figure~\ref{fig:implementation_scenario_ids_nat}. 

 	The specific changes made in the network and \textOF{iptables} rules are as follows:
 	
	\begin{verbatim}
		# enabled ipv4 forwarding
		ipv4.forward = 1 # in /etc/sysctl.conf
		
		# to resolve the DNS by the client
		sh -c "echo nameserver 8.8.8.8 >\
		/etc/resolv.conf"	
		
		# for traffic forwarding with NAT
		iptables -t nat -A POSTROUTING \
		-j MASQUERADE	
	\end{verbatim}
	
	The Snort IDS listens to the Ethernet interface (Eth0) connected to the client and generates alerts based on the rules configured in Snort. The Snort IDS module sends the alert messages to the \textOF{log-record module} at the control-plane. We note that the IDS network function silently listens to the interface and generates alert messages based on the forwarding rules configured in snort. It is not involved in making packet forwarding decisions. Therefore, the network performance in IDS scenario is eventually impacted by the NAT network function. We use snort community rules and customized rules for decision making. A sample set of rules used in Snort for IDS is as follows:
	
	\begin{verbatim}
		alert icmp any any -> \
		any any (msg:"ADMIN-ALERT, ICMP traffic \
		detected";sid:1000004;)
		
		alert tcp any any -> \
		$HOME_NET 80 (msg:"Possible HTTP DoS \
		Attack";sid:1000002;)
		
		alert icmp any any -> \
		$HOME_NET 80 (msg:"Dos Attack suspected";\
		sid:1000001;)
		
		alert tcp $EXTERNAL_NET any -> $HOME_NET \
		445 (msg: "Exploit Detected!"; \
		flow: to_server, established; classtype: \
		attempted-admin; priority: 10; \
		sid: 2094284; rev: 2;)
	\end{verbatim}
	
	%--------------------------------------------------------------------------
	\subsection{Scenario 2: Intrusion Prevention System (IPS)}\label{Secsub:scenario_ips_nat}
\begin{figure}[!ht]
		\includegraphics[scale=0.5]{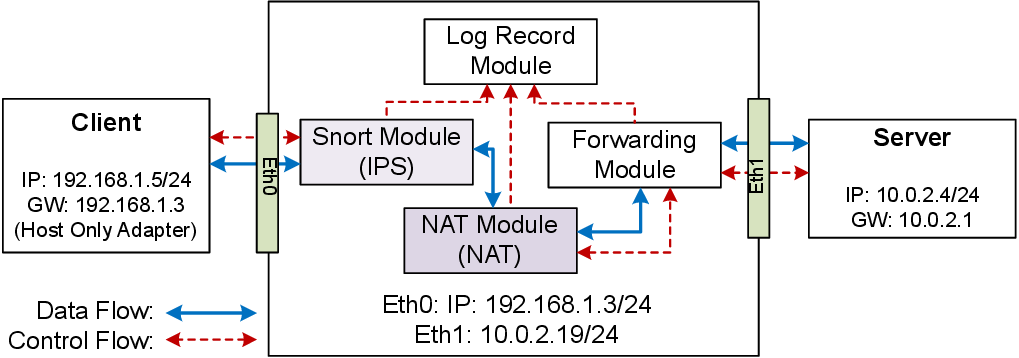}
		\caption{Scenario 2: IPS between client and server}
		\label{fig:implementation_scenario_ips_nat}
	\end{figure}

	Figure~\ref{fig:implementation_scenario_ips_nat} shows the second scenario, where the Snort module acts as an IPS. The Snort module utilizes the NFQ to process all packets sent from the client so that the packets pass through the Snort IPS and the desired action is taken -- whether to \textOF{forward} or \textOF{drop} or \textOF{send alert}. We note that all packets are allowed through the Snort IPS module in this experiment, as our primary objective is to study the impact on network performance. However, any desired action can easily be integrated into the setup. 
	The specific changes made in network and snort configurations in addition to the \textOF{IDS} scenario (refer to Sec~\ref{Secsub:scenario_ids_nat}) are as follows:
	\begin{verbatim}
		# to inspect each packet with NFQUEUE
		iptables -I FORWARD -j NFQUEUE\
		--queue-num=4
		
		# Changes in snort.conf
		config daq: nfq
		config daq_mode: inline
		config daq_var: queue=4
	\end{verbatim}
	
	%======================= Results and Discussion =======================
	\section{Results and Discussion}\label{Sec:Results_discussion}
	We generate TCP and UDP applications using the D-ITG traffic generator~\cite{ditg}. The traffic generator follows the Poisson distribution to generate traffic, in which we vary the average number of packets. Furthermore, each experiment runs 20 times for 30 seconds each. We use the 95\% confidence interval~\cite{hackshawStatisticalFormulaeCalculating2009} to plot the results with a varying average number of packets with payload size 512 bytes. Furthermore, as discussed in Sections~\ref{Secsub:scenario_ids_nat} and \ref{Secsub:scenario_ips_nat}, two scenarios are considered, called \textOF{IDS} and \textOF{IPS}, respectively. The following performance metrics are considered -- average network throughput and jitter for TCP and UDP applications. Whereas latency and the percentage of packet drops are considered for UDP applications as these are the critical parameters for real-time applications. Results obtained using virtual machine-based and containerized deployment are discussed alongside to provide a comparative study on the efficacy of the environments and impact of softwarized security functions on the network performance. 
	
	%\subsection{Results: VM-based Implementation}\label{Secsub:result_vm_all}

	\subsection{Impact on Throughput}\label{Secsub:Result_throughput}
	\subsubsection{TCP Application}\label{Secsubsub:result_throughput_vm}
\begin{figure}[!ht]
		\centering
		
		\subfigure[VMs]{
			\includegraphics[scale=0.45]{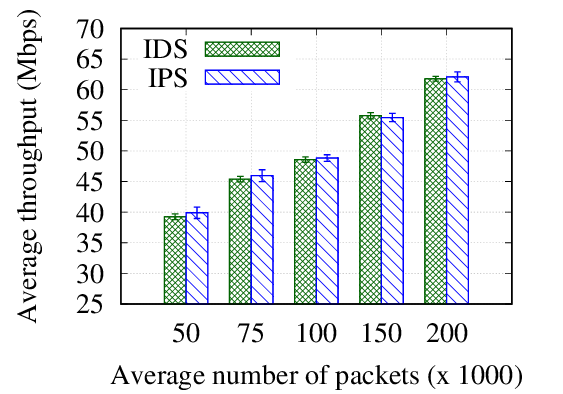}
			\label{fig:result_throughput_tcp_vm}
		}%
		\subfigure[Containers]{
			\includegraphics[scale=0.45]{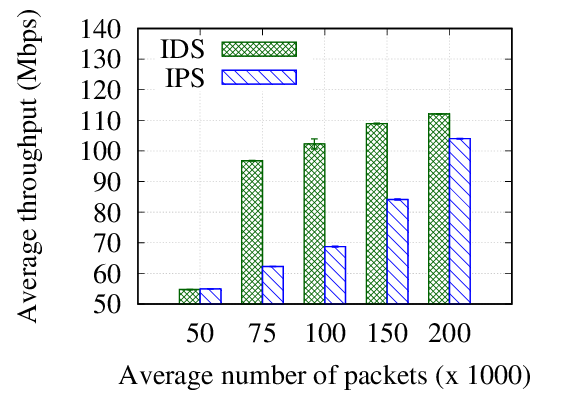}
			\label{fig:result_throughput_tcp_container}
		}		
		\caption{TCP application: Throughput with varying number of packets}
		\label{fig:result_throughput_tcp}
	\end{figure}

	Figure~\ref{fig:result_throughput_tcp} presents the average throughput for TCP applications tested in different network scenarios. We observe that the throughput increases for both \textOF{IPS} and \textOF{IDS} with an increase in the average number of packets. Average throughput for TCP applications is higher in case of containerized environment as compared to VMs. Containers share the host operating system kernel, reducing the overhead compared to VMs. The throughput in \textOF{IDS} remains relatively high as the packet processing is not affected by the IDS module. However, \textOF{IPS} experiences degraded throughput performance. This is because the active threat prevention mechanisms require stateful packet inspection, which is more resource-intensive.
	%We evaluate the impact on throughput with varying packet sizes for both TCP and UDP applications to find out the reasons for the degraded throughput performance by UDP.
	
	%From Figure~\ref{fig:result_throughput_vm}, it is evident that the placement of network security functions on commodity hardware is capable of meeting the data-rate requirements of the emerging applications presented in Table~\ref{table:5G3GPP}. 
	
	\subsubsection{UDP Application}\label{Secsubsub:result_throughput_UDP}
\begin{figure}[!ht]
		\centering
		
		\subfigure[VMs]{
			\includegraphics[scale=0.45]{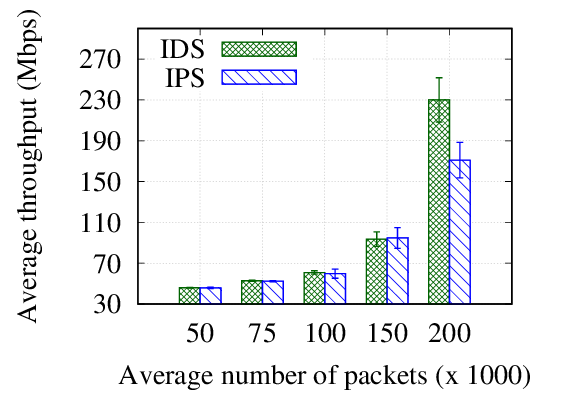}
			\label{fig:result_throughput_udp_vm}
		}%
		\subfigure[Containers]{
			\includegraphics[scale=0.45]{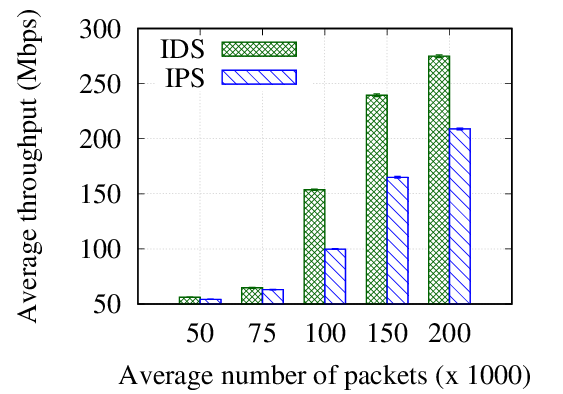}
			\label{fig:result_throughput_udp_container}
		}
		\caption{UDP application: Throughput with varying number of packets}
		\label{fig:result_throughput_UDP}
	\end{figure}

	Figure~\ref{fig:result_throughput_UDP} presents the average
	throughput for UDP applications tested in different scenarios for VMs and containers. We see that the average throughput is degraded in case of \textOF{IPS} when compared with \textOF{IDS}. Furthermore, containerized implementation provides improved throughput as compared to VMs due to  the combination of factors related to resource efficiency and potentially efficient kernel-level packet processing, similar to the results with TCP applications. The average throughput for TCP applications is lower than the UDP applications due to less overhead in UDP. 
	
	\subsection{Impact on Latency}\label{Secsub:Result_latency}

\begin{figure}[!ht]
			\centering
			\subfigure[VMs]{
				\includegraphics[scale=0.45]{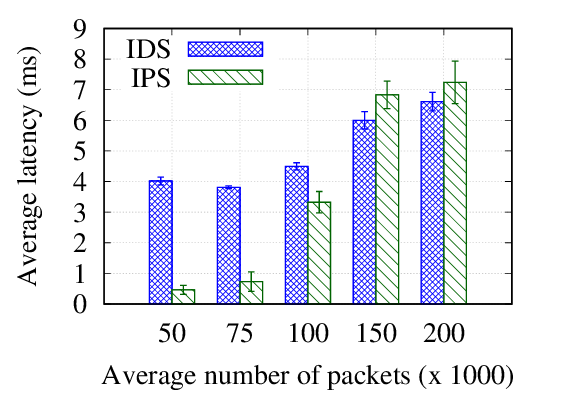}
				\label{fig:result_latency_udp_vm}
			}%
			\subfigure[Containers]{
				\includegraphics[scale=0.45]{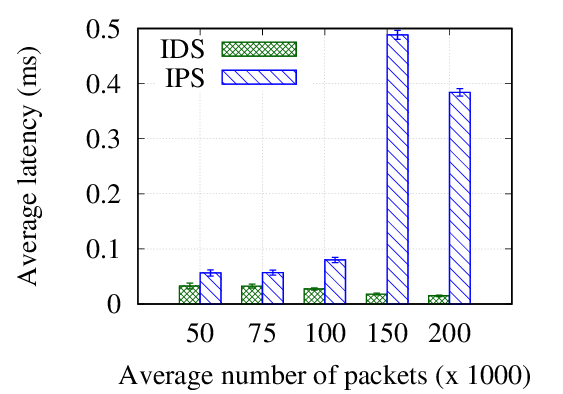}
				\label{fig:result_latency_udp_container}
			}
			\caption{UDP application: Latency with varying number of packets}
			\label{fig:result_latency_udp}
		\end{figure}

		We measure the impact on latency with varying average number of packets in the network. Figure~\ref{fig:result_latency_udp} shows the average latency experienced by packets for both VM-based and containerized implementations with \textOF{IDS} and \textOF{IPS} for UDP applications.
		
		Figure~\ref{fig:result_latency_udp} reveals a pronounced increase in latency with large number of packets in the network. This increased latency is a direct consequence of NAT overloading and packet inspection module. The impact on latency in \textOF{IDS} scenario is less as compared to \textOF{IPS}. This is because of additional packet inspection process in \textOF{IPS} that adds additional delay, which, in turn, produces increased latency in case of \textOF{IPS} as compared to \textOF{IDS}. The \textOF{IDS}'s passive monitoring allows quicker packet delivery, while  \textOF{IPS} introduces delays due to its active intervention even without connection establishment in UDP. However, \textOF{IPS} incurs a lower latency with that of the \textOF{IDS} in the presence of small number of packets. This is because the separate NFQ is assigned for packet processing that does not get overloaded with small number of packets.
		
		Latency observed in case of containers is negligible for smaller number of packets as compared to higher number of packets. However, overall latency for UDP application in case of containers is very less as compared to VMs. This is attributable to combination of better packet processing and efficient resource utilization in containerized environment (as compared to VMs) in addition to the connection-less feature of UDP.

	\subsection{Impact on Jitter}\label{Secsub:Result_jitter}
	
	\subsubsection{TCP Application}\label{Secsubsub:result_jitter_vm}
	
\begin{figure}[!ht]
		\centering
		\subfigure[VMs]{
			\includegraphics[scale=0.45]{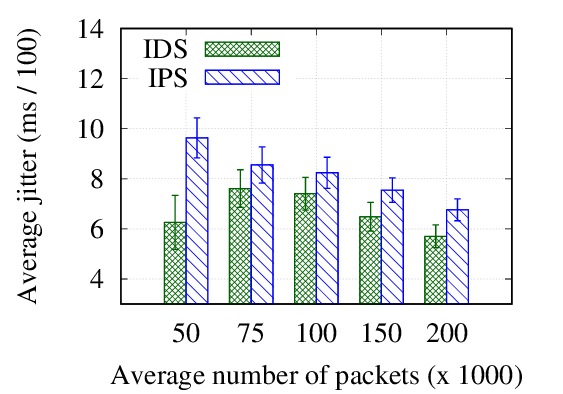}
			\label{fig:result_jitter_tcp_vm}
		}%
		\subfigure[Containers]{
			\includegraphics[scale=0.45]{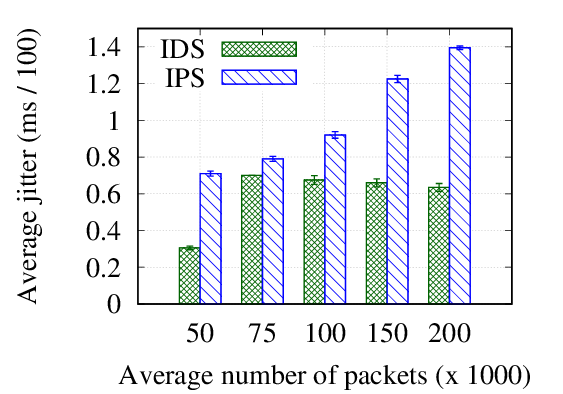}
			\label{fig:result_jitter_tcp_container}
			
		}
		\caption{TCP application: Jitter with varying number of packets}
		\label{fig:result_jitter_tcp}
	\end{figure}

	We measure the jitter with TCP applications, as it is important for many live applications with stringent reliability requirements, such as remote healthcare. Figure~\ref{fig:result_jitter_tcp} shows the average jitter experienced by packets for TCP applications. \textOF{IPS} processing variability leads to higher jitter due to the dynamic analysis of each packet. Impact of jitter was more in case of VMs than in containers similar to the throughput and latency. 
	
	\subsubsection{UDP Application}\label{Secsubsub:result_jitter_container}
	
	\begin{figure}[!ht]
		\centering
		\subfigure[VMs]{
			\includegraphics[scale=0.45]{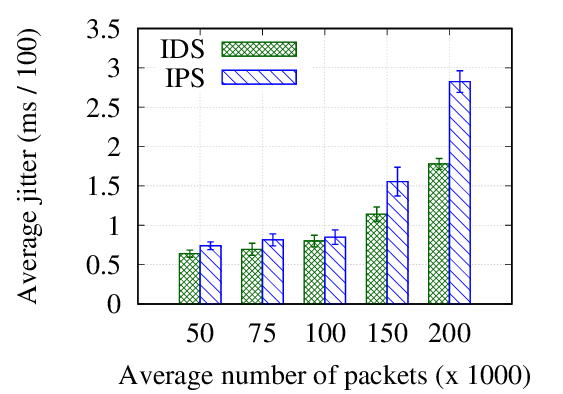}
			\label{fig:result_jitter_udp_vm}
		}%
		\subfigure[Containers]{
			\includegraphics[scale=0.45]{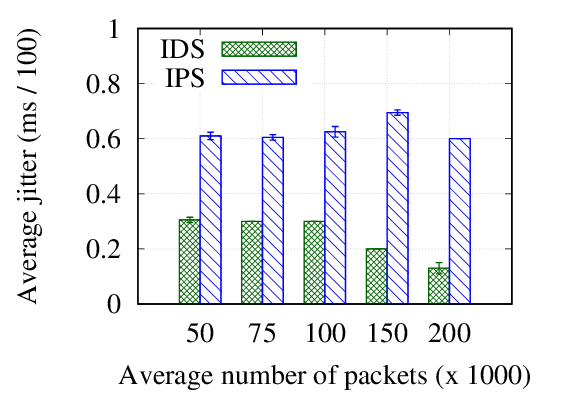}
			\label{fig:result_jitter_udp_container}
		}
		\caption{UDP application: Jitter with varying number of packets }
		\label{fig:result_jitter_udp}
	\end{figure}
	
	Jitter is also important for many fault-tolerant live applications, such as audio-video conferencing. An increased jitter is observed under \textOF{IPS} conditions compared to \textOF{IDS} for UDP applications, as shown in Figure~\ref {fig:result_jitter_udp}. This can be attributed to the variable and intensive packet analysis required by IPS functions, introducing fluctuations in handling times. In contrast, \textOF{IDS} processing remains relatively uniform, leading to a lower jitter. While UDP's lack of acknowledgment mechanisms mitigates the impact on jitter under \textOF{IDS} but does not fully shield it from the interruption and inspection processes of \textOF{IPS}, hence, increased jitter is observed. This emphasizes the non-negligible impact of software-based security functions on jitter, particularly for real-time applications. Similar pattern is observed for containers with respect to jitter in \textOF{IDS} and \textOF{IPS} scenarios. Furthermore, VMs experience higher jitter compared to containers. This is due to the additional layer of abstraction and resource constraints associated with virtualization.

	\subsection{Impact on Packet Drop}\label{Secsub:Result_packet_drop}
\begin{figure}[!ht]
		\centering
		\subfigure[VMs]{
			\includegraphics[scale=0.45]{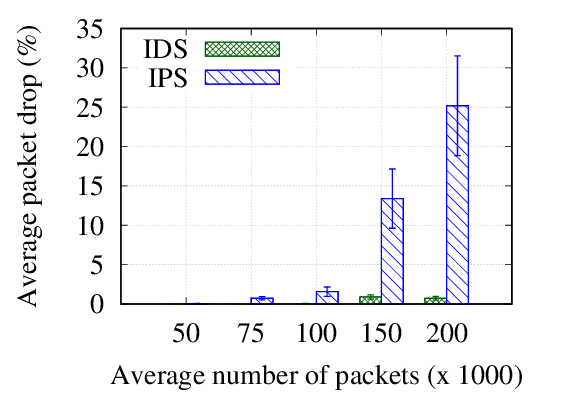}
			
			\label{fig:result_packet_drop_vm}
		}%
		\subfigure[Containers]{
			\includegraphics[scale=0.45]{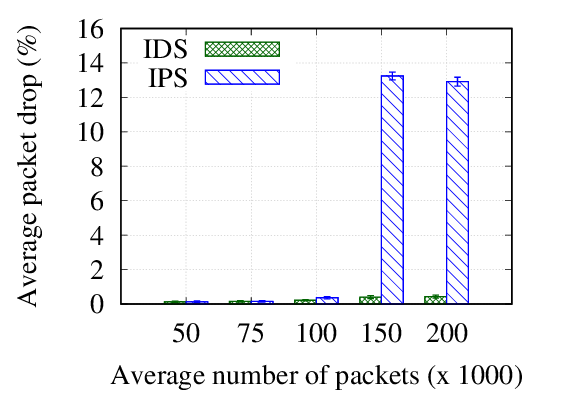}
			
			\label{fig:result_packet_drop_container}
		}
		\caption{Percentage of packet drops for UDP applications}
		\label{fig:result_packetdrop_udp}
	\end{figure}

	We present the percentage of packet drops in UDP applications with different security scenarios for VMs and containers in Figure~\ref{fig:result_packetdrop_udp}. Similar to the throughput, latency, and jitter, we see a degraded network performance in terms of packet loss in the presence of softwarized security functions.  
	
	The percentage of packet-drop is higher in case of \textOF{IPS} than that of the \textOF{IDS}. Furthermore, the percentage of packet drop increases with increasing number of packets in the network similar to the latency. This is due to the overloading of the IPS and NAT network functions with large number of packets.	
	
	Figure~\ref{fig:result_packet_drop_container} presents the percentage of packet drops in the containerized environment. We see the similar performance trends as in VMs with a reduced percentage of packet drop.  	
	Considering the reliability requirements presented in Table~\ref{table:5G3GPP} and packet drop with UDP applications in Figures~\ref{fig:result_packet_drop_vm} and \ref{fig:result_packet_drop_container}, TCP should be used for achieving high reliability of the underlying use-case scenarios. However, UDP can be used for increased throughput with low-reliability requirements.			

	In summary, we see that the softwarized security functions are capable of achieving the diverse QoS requirements in terms of throughput and latency for applications with moderate latency requirements. However, it may not be suitable for ultra-low latency requirements. Furthermore, redundant placement of security functions is also required to achieve high reliability. We note that the network performances may vary depending on the commodity hardware configuration. Therefore, the performance of the softwarized security functions and QoS requirements of associated applications must be taken into consideration while placing them on commodity hardware. We also understand that performance of security functions in terms of impact on QoS parameters is found to be less in containerized environment as compared to VM-based implementation.
	%======================================================
	
	\section{Use-Case Scenario: 5G Core Network}\label{Sec:Use_case_5G}
\begin{figure}[!ht]
		\centering
		\includegraphics[scale=1.1]{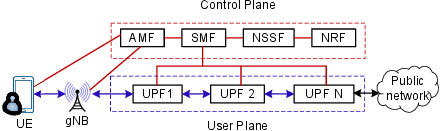}
		\caption{Standalone architecture of 5G network}
		\label{fig:Architecture_5g}
	\end{figure}

	To study the network performance in a real-world use-case scenario, we consider the standalone architecture of 5G network, as presented in Figure~\ref{fig:Architecture_5g}. 
	We deploy the 5G core network using Open5GS~(\url{https://open5gs.org/open5gs/}) open-source software platform. The 5G RAN is deployed using UERANSIM~(\url{https://github.com/aligungr/UERANSIM}) open-source software platform. UERANSIM and Open5GS are integrated together to have an end-to-end softwarized 5G network.  We setup the experiment platform in a host machine with the similar configuration as mentioned in Section~\ref{Sec:Implementation}. We create three guest machines, one for UE and gNB placement, another for 5G core network placement with security functions, and the other acts as a server to the UE, similar to the setup explained in Section~\ref{Sec:Implementation}. The components inside IDS and IPS modules are similar as discussed in Section~\ref{Sec:Implementation}.
	\begin{figure}[!ht]
		\centering
		\subfigure[5G UPF as IDS]{
			\includegraphics[scale=1.0]{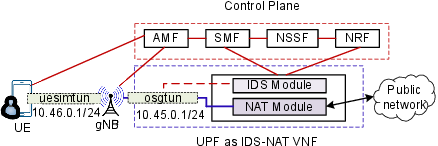}
			\label{fig:5g_ids}
		}
		\subfigure[5G UPF as IPS]{
			\includegraphics[scale=1.0]{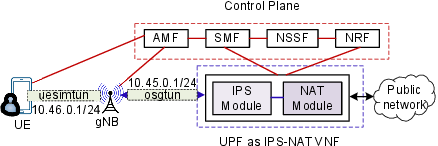}
			\label{fig:5g_ips}
		}
		\caption{5G UPF as IDS and IPS}
		\label{fig:5g_ids_ips}
	\end{figure}

	We configure the UPF so that it acts as either IDS or IPS along with NAT. Similar to the generalized setup of the network, NAT functionality is also an integral part of the 5G network system. We consider the same configurations scenarios for IDS and IPS, as described in Section~\ref{Sec:Implementation}.
	
	\subsection{Results and Discussion}\label{Secsub:Results_use_case_5g}
	We present the results on network performance in terms of throughput, latency, and packet drop similar to Section~\ref{Sec:Results_discussion}.
	
\begin{figure}[!ht]
		\centering
		\subfigure[TCP application]{
			\includegraphics[scale=0.46]{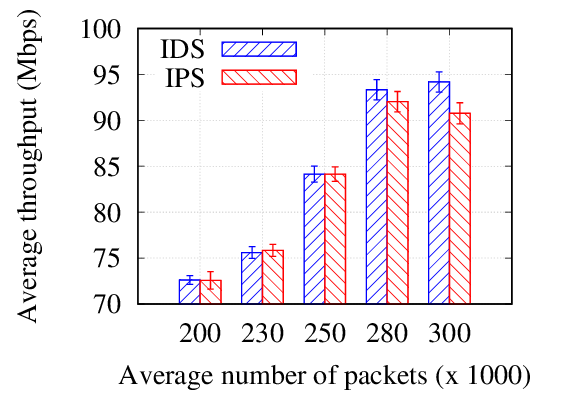}
			\label{fig:result_throughput_tcp_5g}
		}%
		\subfigure[UDP application]{
			\includegraphics[scale=0.46]{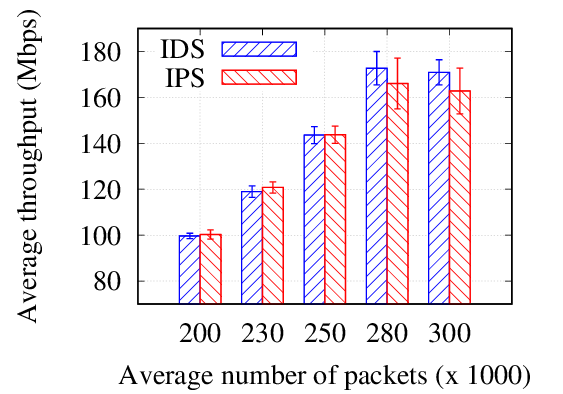}
			\label{fig:result_throughput_udp_5g}
		}
		\caption{Throughput with varying number of packets}
		\label{fig:result_throughput_5g}
	\end{figure}

	Figure~\ref{fig:result_throughput_5g} presents the average network throughput with varying number of packets for TCP and UDP applications. We see that the average throughput increases linearly initially with less number of packets for both TCP and UDP applications. However, it gets saturated for large number of packets due to the network capacity constraint. As expected, the UDP provides higher throughput than the TCP, as shown in Figures~\ref{fig:result_throughput_tcp_5g} and \ref{fig:result_throughput_udp_5g}. Furthermore, we see that \textOF{IDS} and \textOF{IPS} yield a similar throughput for small number of packets. However, \textOF{IDS} yields a higher throughput than \textOF{IPS} with large number of packets. This is because the packets are forwarded without sending them to the NFQ module in \textOF{IDS}. The Snort module only listens to the 5G tunnel interface and sends alerts based on the rule-set. In contrast, the packets are always sent to the NFQ module and analyzed before being forwarded/dropped, when the security function works as \textOF{IPS}. Therefore, the queue is overloaded in the presence of large number of packets. Hence, packets get dropped, which, in turn, leads to decreased network throughput.
	
	\subsubsection{Latency}\label{Secsubsub:Result_latency_5g}
\begin{figure}[!ht]
		\centering
		\subfigure[TCP application]{
			\includegraphics[scale=0.46]{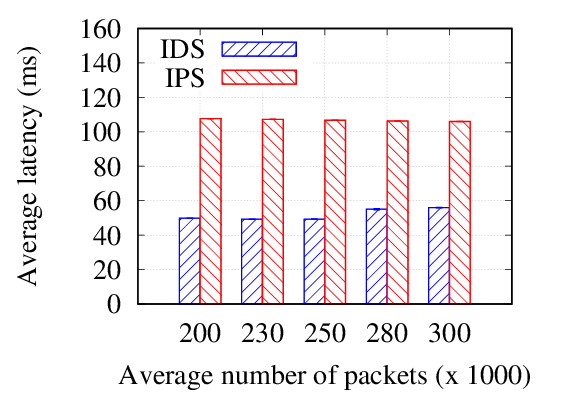}
			\label{fig:result_delay_tcp_5g}
		}%
		\subfigure[UDP application]{
			\includegraphics[scale=0.46]{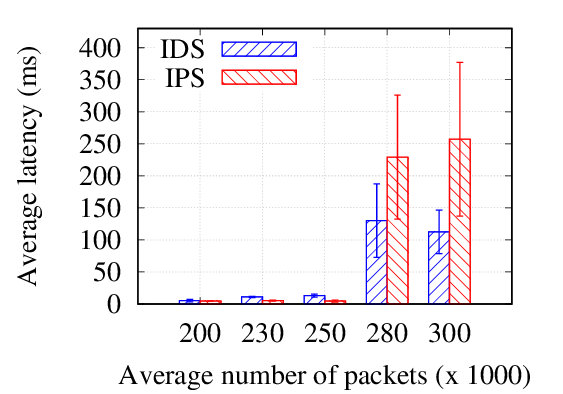}
			\label{fig:result_delay_udp_5g}
		}
		\caption{Latency with varying number of packets}
		\label{fig:result_delay_5g}
	\end{figure}

	Figures~\ref{fig:result_delay_5g} presents the network latency with varying number of packets. Similar to the throughput, the latency experienced by TCP applications (refer to Figure~\ref{fig:result_delay_tcp_5g}) is more than the UDP applications (refer to Figure~\ref{fig:result_delay_udp_5g}). This is because TCP is connection-oriented, which incurs additional delay compared to UDP. Furthermore, the latency experienced by TCP applications is almost constant irrespective of the number of packets in the network, as presented in Figure~\ref{fig:result_delay_tcp_5g}. This is due to the fact that TCP adaptively adjusts the congestion-window size, which eventually affects the throughput, as presented in Figure~\ref{fig:result_throughput_5g}. On the other hand, the latency experienced by UDP applications is very low for small number of packets. However, the latency increases non-linearly with large number of packets due to congestion at the NFQ module. This also leads to packet drop, which is discussed in the next section.

	\subsubsection{Packet Drop}\label{Secsubsub:Result_packet_drop_5g}
\begin{figure}[!ht]
		\centering
		\includegraphics[scale=0.46]{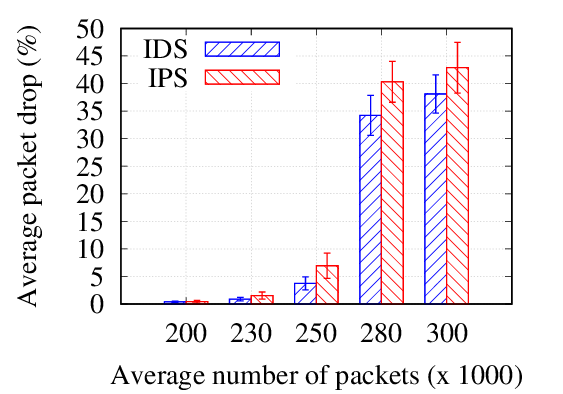}
		\caption{Percentage of packet drop with varying number of packets for UDP application}
		\label{fig:result_packet_drop_5g}
	\end{figure}

	Figure~\ref{fig:result_packet_drop_5g} presents the percentage of packet drop for UDP applications with varying number of packets in the network. We see that the percentage of packet drop is very low in the presence of small number of packets for both \textOF{IDS} and \textOF{IPS}. However, it increases non-linearly with large number of packets. This is because of the NAT and NFQ overloading, which causes increased packet drop. Furthermore, the percentage of packet drop is higher in \textOF{IPS} than \textOF{IDS} due to the NFQ overflow.
	
	In summary, we observe a similar network performance in 5G network enabled with the softwarized security functions.

	%====================================================
	\section{Conclusion and Implications to Future Work}\label{Sec:Conclusion}
	We studied the performance of network security functions when placed on commodity hardware. We provided thorough implementation details of a secure softwarized network with IDS and IPS. The impact on the network performance is studied in the presence of TCP and UDP applications for VM-based and containerized implementation. Extensive experiment results showed that softwarized network functions significantly impact the network performance in terms of throughput, latency, jitter, and packet drop. Furthermore, it is observed that the containerized placement offers improved performance than VM-based implementation. We also applied the implementation in a 5G network as a use-case scenario. We observed the similar pattern in the results on throughput, latency, and packet drop.
	
	We identify some future research implications as more study is required to understand different aspects of softwarized networks.
	
	\textbf{Platform to place software functions:} As evident from this study and the existing works, the hardware and software platforms, on which the softwarized functions are placed, also affect the network performance. Therefore, further study is required with different hardware and software platforms to have \textit{optimized} softwarized network designed for application-specific service provisioning.
	
	\textbf{Software tools to create network security functions:} In this work, we used Snort to create IDS and IPS security functions and analyzed the impact on network performance. Other software platforms, such as Suricata~(\url{https://suricata.io/}), can also be used to conduct a comparative study. We note that a few works presented a comparative study between the Snort and Suricata to create the IDS. However, they mainly focused on the efficiency of the platform without considering the impact on the network performances.
	
	\textbf{Use of TCP and UDP applications:} We observed a high throughput and high percentage of packet drop for UDP applications when a large number of packets are generated in the network. On the other hand, TCP yields lower throughput, high latency, and reliable packet delivery. Furthermore, the use of TCP and UDP not only depends on the underlying applications, but also on the network setup. Consequently, in-depth study on the trade-offs between packet-size, average number of packets generated by an application, and the protocol in-use (UDP or TCP) is required for QoS-guaranteed service provisioning.
	
	\textbf{Application-specific network slicing:} With the introduction of network slicing~\cite{afolabiNetworkSlicingSoftwarization2018}, we can create multiple logical networks, each serving different applications based on their service-level agreements. Therefore, in addition to the security functions listed in this work, the performance of each slice should be studied in the presence of different user plane functions such as packet routing and forwarding, policy enforcement, video optimizer, and network monitoring. In such a case, studying the end-to-end network performance of each slice is required.
	
	%=====================================================
	
	\bibliographystyle{IEEEtran}
	\bibliography{Snort_IDS_IPS}

\end{document}